# Carrier-controlled ferromagnetism in SrTiO$_3$

Pouya Moetakef[1,*], James R. Williams[2,*], Daniel G. Ouellette[3], Adam Kajdos[1], David Goldhaber-Gordon[2], S. James Allen[3], and Susanne Stemmer[1,**]

[1]Materials Department, University of California, Santa Barbara, California, 93106-5050, USA

[2]Department of Physics, Stanford University, Stanford, California, 94305-4045, USA

[3]Department of Physics, University of California, Santa Barbara, California, 93106-9530, USA

* These authors contributed equally to the study.

** Corresponding author.  Email: stemmer@mrl.ucsb.edu

**Abstract**

Magnetotransport and superconducting properties are investigated for uniformly La-doped SrTiO$_3$ films and GdTiO$_3$/SrTiO$_3$ heterostructures, respectively. GdTiO$_3$/SrTiO$_3$ interfaces exhibit a high-density two-dimensional electron gas on the SrTiO$_3$-side of the interface, while for the SrTiO$_3$ films carriers are provided by the dopant atoms. Both types of samples exhibit ferromagnetism at low temperatures, as evidenced by a hysteresis in the magnetoresistance. For the uniformly doped SrTiO$_3$ films, the Curie temperature is found to increase with doping and to coexist with superconductivity for carrier concentrations on the high-density side of the superconducting dome. The Curie temperature of the GdTiO$_3$/SrTiO$_3$ heterostructures scales with the thickness of the SrTiO$_3$ quantum well. The results are used to construct a stability diagram for the ferromagnetic and superconducting phases of SrTiO$_3$.

Two-dimensional electron liquids at interfaces between two insulating oxides, such as LaAlO$_3$/SrTiO$_3$, have generated tremendous excitement because of unique properties, such as strong electron correlations, superconductivity and ferromagnetism [1-4]. While superconductivity is a known bulk property of SrTiO$_3$ [5], ferromagnetism is not a property of either material and has so far only been observed for LaAlO$_3$/SrTiO$_3$ interfaces [3,6-9]. Furthermore, recent studies indicate the coexistence of superconductivity and ferromagnetism at these interfaces [6-8]. These observations raise a number of scientific questions. Among these, the most important question is whether ferromagnetism, and its coexistence with a superconducting state, is an intrinsic, many-body property of a two-dimensional oxide electron liquid in SrTiO$_3$, or caused by extrinsic contributions, such as disorder, defects, or impurities [10]. Pointing to the latter are observations of ferromagnetic regions that form non-uniform patches [8], and inconsistencies in the appearance of magnetism in samples grown under the same conditions [11]. Furthermore, a large fraction of the interfacial charge at LaAlO$_3$/SrTiO$_3$ interfaces appears to be localized, indicating the presence of traps [12], and mobile charge densities are an order of magnitude less than what is needed to compensate for the polarization discontinuity [13].

In this study, ferromagnetism in SrTiO$_3$, and its coexistence with superconductivity, are investigated in two different structures: extreme-electron-density electron liquids at GdTiO$_3$/SrTiO$_3$ interfaces and uniformly doped SrTiO$_3$ films. The charge carriers, which reside in the SrTiO$_3$ for both structures, have fundamentally different origins: for the uniformly doped SrTiO$_3$ films, La dopant atoms provide the carriers, while for the GdTiO$_3$/SrTiO$_3$ interfaces carriers are due to the interface. In particular, similar to LaAlO$_3$/SrTiO$_3$ interfaces, GdTiO$_3$/SrTiO$_3$ heterostructures exhibit a fixed polar charge, due to the dipole moment associated with the [$Gd^{3+}O^{2-}$/$Ti^{3+}O_2^{4-}$] bilayers along the (001) surface normal of $GdTiO_3$. Unlike $LaAlO_3$/$SrTiO_3$, however, $GdTiO_3$/$SrTiO_3$ interfaces grown by molecular beam epitaxy (MBE) exhibit mobile carrier densities that are remarkably well predicted by the electrostatic requirements of the compensation of the polar discontinuity at the interface [14]. Carrier densities are 3-4×$10^{14}$ cm$^{-2}$, or ~ ½ electron per surface unit cell, *at each $GdTiO_3$/$SrTiO_3$ interface* for all heterostructures containing more than one unit cell of $SrTiO_3$ [14]. The staggered band alignment causes the mobile charge to reside in the $SrTiO_3$, which is the wider band gap material [14]. These interfaces allow for doping of narrow $SrTiO_3$ quantum wells to extremely large and mobile carrier concentrations. For example, by sandwiching a few-unit-cell-thick $SrTiO_3$ layer between two $GdTiO_3$ layers, carrier concentrations in the $SrTiO_3$ approach one electron per site, conditions under which electron correlation effects should appear [15,16].

Here we show that both types of samples, uniformly doped $SrTiO_3$ films and high-density two-dimensional electron gases at $GdTiO_3$/$SrTiO_3$ interfaces, exhibit ferromagnetism. A ferromagnetic phase stability region appears on the high-density side of the superconducting dome in the phase diagram, and, at an intermediate carrier concentration, both phases coexist.

Three types of high-electron-concentration samples grown by MBE were investigated here: $GdTiO_3$/$SrTiO_3$ heterostructures grown on $(LaAlO_3)_{0.3}(Sr_2AlTaO_6)_{0.7}$ (LSAT) substrates, either containing a single electrically active interface ($GdTiO_3$/$SrTiO_3$/LSAT) or two active interfaces ($GdTiO_3$/$SrTiO_3$/$GdTiO_3$/LSAT), and two uniformly La-doped $SrTiO_3$ films of about 40 nm thickness (nominal dopant densities of $1 \times 10^{20}$ cm$^{-3}$ and $9 \times 10^{20}$ cm$^{-3}$, respectively) grown on LSAT substrates. Details of the MBE growth procedures have been described in detail elsewhere [14,17,18]. Previous studies have demonstrated a close correspondence between La dopant concentrations in the $SrTiO_3$ and mobile carrier densities, down to dopant concentrations of $\sim 1\times10^{17}$ cm$^{-3}$ [19]. This provides evidence for the absence of additional sources of carriers, such as oxygen vacancies, and of traps such as transition metal impurities or Sr vacancies (which would reduce the carrier concentrations relative to the La dopant concentration), on the ppm level. Furthermore, positron annihilation studies of the La-doped films indicate Sr vacancy concentrations of no more than $\sim 1\times10^{16}$ cm$^{-3}$ in these films [20]. We note that the carrier densities in the uniformly doped samples investigate here are slightly less than the dopant concentration, because of the well-known surface depletion of SrTiO$_3$ [21]. For the SrTiO$_3$/GdTiO$_3$ heterostructures, the thickness of the SrTiO$_3$ layer was varied between 1 and 10 nm.

Ohmic contacts to the GdTiO$_3$ and SrTiO$_3$ layers were 300 nm Au/50 nm Ti and 300 nm Au/20 nm Ni/40 nm Al, respectively, with Au contact being the top-most layer in each case. Magnetoresistance measurements were carried out in Van der Pauw geometry using a Physical Properties Measurement System (Quantum Design PPMS). The sweep rates were varied between 0.005 and 0.01 T/s and all results (such as hysteresis and Curie temperature) shown here were found to be independent of the sweep rate. For the Hall measurements the magnetic field $B$ was varied between $\pm 1$ T. The two-dimensional (2D) carrier concentration $n$ was calculated as $n = 1/(e \cdot R_H)$ where $R_H$ is the measured 2D Hall coefficient. Low temperature measurements of the uniformly doped samples were performed in a dilution fridge, capable of temperatures down to 12 mK and magnetic field up to 9 T. The low temperature magnetoresistance was measured using a standard lock-in technique at a frequency of 13 Hz. Magnetization was measured using a SQUID magnetometer (Quantum Design Magnetic Properties Measurement System).

Figure 1(a) shows the longitudinal magnetoresistance, $R(B)$, where $B$ is the magnetic field, of a 1-nm SrTiO$_3$ quantum well sandwiched between two GdTiO$_3$ layers, measured in four-terminal geometry. The sheet carrier density of this sample, as extracted from the Hall resistance, is $\sim 9\times10^{14}$ cm$^{-2}$ at room temperature and $\sim 1\times10^{15}$ cm$^{-2}$ at 5 K, corresponding to a three-dimensional (3D) carrier density of $\sim 1\times10^{22}$ cm$^{-3}$ in the 1-nm quantum well. Negative magnetoresistance is apparent at low temperatures (below 10 K), as has been previously reported for La-doped SrTiO$_3$ [22]. Concurrent with the appearance of negative magnetoresistance, $R(B)$ becomes hysteretic, indicating ferromagnetism, when the magnetic field was in-plane and perpendicular to the current. We use the appearance of hysteresis to define the Curie temperature of the ferromagnetic transition, $T_{Cm}$. We note that although extrinsic effects, the possibility of which is further discussed below, in ferromagnets may give rise to magnetoresistance hysteresis, the presence of a hysteresis nonetheless implies ferromagnetic ordering [23]. The shape of the hysteresis is very similar to that observed by Brinkman et al. for a LaAlO$_3$/SrTiO$_3$ interface [3]; however, here it occurs at almost an order-of-magnitude higher temperature than in ref. [3].

Heterostructures with a single interface and 1-nm SrTiO$_3$, which contain half the carrier density or less (sheet carrier densities of 4.6×10$^{14}$ cm$^{-2}$ at room temperature and 1.9×10$^{14}$ cm$^{-2}$ at 5 K; the latter corresponding to a 3D concentration of 1.9×10$^{21}$ cm$^{-3}$), also show ferromagnetic hysteresis, albeit at a lower $T_{Cm}$ of about 3 K [Fig. 1(b)]. Samples with thicker (2 nm) SrTiO$_3$ layers did not show any magnetoresistance hysteresis above 2 K. In addition to the hysteresis in $R(B)$, all samples that exhibit hysteresis show an increase in resistance as the temperature is lowered, indicating a departure in the nature of conduction from conventional metallic systems [24]. For data from a sample with a slightly thinner SrTiO$_3$ layer, see ref. [24].

GdTiO$_3$ layers are ferrimagnetic [17,25] and it is therefore entirely conceivable that the magnetism of the carriers in the SrTiO$_3$ is induced by the proximity to the GdTiO$_3$. However, the following observations can also be made: (i) ferromagnetism is absent in samples with SrTiO$_3$ layers that are thicker than 2 nm, despite similar sheet carrier densities and proximity to the GdTiO$_3$; (ii) the Curie temperature of GdTiO$_3$ is substantially higher (~ 20 K, see Fig. 2) than $T_{Cm}$ of any of the two-dimensional electron liquids investigated here; and (iii) ferromagnetism also appears in highly-doped bulk SrTiO$_3$, when no GdTiO$_3$ layer is present, albeit at much lower temperatures, as will be discussed next.

Figure 3 shows $R(B)$ for 44 nm-thick SrTiO$_3$, uniformly doped with La to a density of $9\times10^{20}$ cm$^{-3}$, measured at $T = 12$ mK. Here, $B$ is applied perpendicular to the current and parallel to the growth direction (i.e. perpendicular to the (001) growth surface). Magnetoresistance hysteresis is evident for this sample. The hysteresis shape is very similar to that of the two-dimensional electron gases discussed above and to that at a LaAlO$_3$/SrTiO$_3$ interface [3]. The onset of the hysteresis occurs at $T_{Cm} = 700$ mK, substantially lower than that of the two-dimensional electron liquid of the GdTiO$_3$/SrTiO$_3$ samples. The carrier concentration is above the maximum density for which SrTiO$_3$ becomes superconducting [26]; hence only the ferromagnetic behavior is evident. La-doped SrTiO$_3$ samples with a reduced doping density of $1\times10^{20}$ cm$^{-3}$ (Fig. 4) lie in a regime where superconducting correlations are expected to be important [22,26]. In Fig. 4, evidence for both a superconducting-zero-resistance-state (defined by a transition temperature, $T_{Cs}$) and a ferromagnetic state is apparent. At 12 mK, sweeping B from negative to positive fields, two features indicative of the two states are measured: a sharp drop to $R(B) = 0$ below $B = -1.7$ T, followed by a departure from $R(B) = 0$ at $B = 0.2$ T (red curve in Fig. 4a). This departure from the zero-resistance state is hysteretic (Fig. 4b). The temperature dependence of $R(B)$ (Fig. 4a) shows the evolution of the superconducting and ferromagnetic states and demonstrates that the two transition temperatures are different ($T_{Cs}$ = 315 mK and $T_{Cm}$ = 500 mK for this sample). At very low temperatures the superconducting ground state dominates and only a weak hysteresis is observed. As the temperature is increased, a slow transition to a ferromagnetic ground state is apparent. At 450 mK, only the hysteretic component of $R(B)$ remains. The magnetoresistance hysteresis of the La-doped $SrTiO_3$ samples shows that ferromagnetism occurs in $SrTiO_3$ even in the absence of any proximity to a ferrimagnetic layer and is thus a bulk property of $SrTiO_3$ at sufficiently high La doping concentrations. A third sample with an even lower La doping concentration ($3 \times 10^{19}$ cm$^{-3}$ grown on a $SrTiO_3$ substrate) only shows superconductivity below 100 mK and no ferromagnetism. Bulk doping of $SrTiO_3$ with La to higher carrier densities is possible [16,27], but to obtain the carrier densities of the quantum wells (up to $1 \times 10^{22}$ cm$^{-3}$ in this study), the material would be more appropriately described as Sr-doped $LaTiO_3$, which is a well-known antiferromagnet [28,29]. $LaTiO_3$, up to Sr-doping concentration of ~10%, exhibits a weak, canted ferromagnetic moment [28]. Although even the highest La-doping concentration (~ 5% La doping) in this study is far from this regime, it is possible that locally such weak ferromagnetic correlations exist.

Figure 5 summarizes the correlation between the 3D carrier concentration in $SrTiO_3$ and the ferromagnetic (this study) and superconducting transition temperatures (literature and this study). Data from both bulk and two-dimensional electron liquids is shown in the same graph. For the uniformly doped films, a ferromagnetic phase stability region appears on the high-density side of the superconducting dome in the phase diagram, and, for a range of carrier concentrations, both phases coexist, as was evident in Fig. 4. The results from these samples provide evidence that ferromagnetic phase of $SrTiO_3$ at sufficiently high La dopant concentrations is likely *not* mediated by unintentional defects such as oxygen vacancies or Fe impurities. Impurity-related ferromagnetism typically persists to very high temperatures (see e.g., ref. [10]), which is different from what is observed here. The results from the uniformly doped films are consistent with recent reports of a field-induced Kondo effect in field-gated $SrTiO_3$ structures, which can be considered a precursor to the ferromagnetism at higher carrier densities [30]. When the data from the heterostructures and uniformly doped films are combined, as shown in Fig. 5, the ferromagnetism appears to depend strongly carrier concentration, even though carriers are introduced via two fundamentally different mechanisms (interface and conventional doping). However, more studies are needed to determine the degree to which the proximity to one or two $GdTiO_3$ interfaces, respectively, acts to increase the ferromagnetic transition temperature in the heterostructures, and to clarify the role of the La-dopants in the uniformly doped films.

Theoretical models of the ferromagnetism in doped $SrTiO_3$ should address the potential roles of band filling, Coulomb repulsion [31] and the coexistence with a superconducting state. Theory should consider the importance of the role of the degree of filling of different bands and subbands in the $SrTiO_3$ [6], which are derived from the three degenerate $t_{2g}$ *d*-orbitals of Ti [32], the role of strain and band degeneracy (note that all samples that exhibited ferromagnetism were grown on LSAT, causing compressive biaxial stress in the $SrTiO_3$) as well as the shape of the observed hysteresis, in particular the decrease in resistance as the magnetic field reaches the coercive field. Finally, the dependence of ferromagnetism on the quantum well thickness in the $SrTiO_3$ heterostructures implies that electric field gating may be used to modulate both ferromagnetism and superconductivity.

The authors thank Leon Balents and Allan MacDonald for many useful discussions. P. M. thanks Tyler Cain and Clayton Jackson for help with the thin film growth experiments. S.S. acknowledges support by the DOE Office of Basic Energy Sciences (Grant No. DEFG02-02ER45994). P.M was supported by the U.S. National Science Foundation (Grant No. DMR-1006640). J. R. W. and D. G-G. acknowledge support by a MURI program of the Army Research Office (Grant No. W911-NF-09-1-0398) and D. G. O and S. J. A. by the MRSEC Program of the National Science Foundation under Award No. DMR 1121053.

# References

[1] A. Ohtomo, D. A. Muller, J. L. Grazul, and H. Y. Hwang, *Artificial charge-modulation in atomic-scale perovskite titanate superlattices*, Nature **419**, 378-380 (2002).

[2] A. Ohtomo and H. Y. Hwang, *A high-mobility electron gas at the LaAlO$_3$/SrTiO$_3$ heterointerface*, Nature **427**, 423-426 (2004).

[3] A. Brinkman, M. Huijben, M. Van Zalk, J. Huijben, U. Zeitler, J. C. Maan, W. G. Van der Wiel, G. Rijnders, D. H. A. Blank, and H. Hilgenkamp, *Magnetic effects at the interface between non-magnetic oxides*, Nature Mater. **6**, 493-496 (2007).

[4] N. Reyren, S. Thiel, A. D. Caviglia, L. F. Kourkoutis, G. Hammerl, C. Richter, C. W. Schneider, T. Kopp, A. S. Ruetschi, D. Jaccard, M. Gabay, D. A. Muller, J. M. Triscone, and J. Mannhart, *Superconducting interfaces between insulating oxides*, Science **317**, 1196-1199 (2007).

[5] J. F. Schooley, W. R. Hosler, and M. L. Cohen, *Superconductivity in semiconducting SrTiO$_3$*, Phys. Rev. Lett. **12**, 474-475 (1964).

[6] D. A. Dikin, M. Mehta, C. W. Bark, C. M. Folkman, C. B. Eom, and V. Chandrasekhar, *Coexistence of Superconductivity and Ferromagnetism in Two Dimensions*, Phys. Rev. Lett. **107**, 056802 (2011).

[7] L. Li, C. Richter, J. Mannhart, and R. C. Ashoori, *Coexistence of magnetic order and two-dimensional superconductivity at LaAlO$_3$/SrTiO$_3$ interfaces*, Nat. Phys. **7**, 762-766 (2011).

[8] J. A. Bert, B. Kalisky, C. Bell, M. Kim, Y. Hikita, H. Y. Hwang, and K. A. Moler, *Direct imaging of the coexistence of ferromagnetism and superconductivity at the LaAlO$_3$/SrTiO$_3$ interface*, Nat. Phys. **7**, 767-771 (2011).

[9] A. J. Millis, *Oxide Interfaces: Moment of magnetism*, Nat. Phys. **7**, 749-750 (2011).

[10] D. A. Crandles, B. DesRoches, and F. S. Razavi, *A search for defect related ferromagnetism in SrTiO$_3$*, J. Appl. Phys. **108**, 053908 (2010).

[11] M. R. Fitzsimmons, N. W. Hengartner, S. Singh, M. Zhernenkov, F. Y. Bruno, J. Santamaria, A. Brinkman, M. Huijben, H. J. A. Molegraaf, J. de la Venta, and I. K. Schuller, *Upper Limit to Magnetism in LaAlO$_3$/SrTiO$_3$ Heterostructures*, Phys. Rev. Lett. **107**, 217201 (2011).

[12] M. Sing, G. Berner, K. Goss, A. Muller, A. Ruff, A. Wetscherek, S. Thiel, J. Mannhart, S. A. Pauli, C. W. Schneider, P. R. Willmott, M. Gorgoi, F. Schafers, and R. Claessen, *Profiling the Interface Electron Gas of LaAlO$_3$/SrTiO$_3$ Heterostructures with Hard X-Ray Photoelectron Spectroscopy*, Phys. Rev. Lett. **102**, 176805 (2009).

[13] M. Huijben, A. Brinkman, G. Koster, G. Rijnders, H. Hilgenkamp, and D. H. A. Blank, *Structure-Property Relation of SrTiO$_3$/LaAlO$_3$ Interfaces*, Adv. Mater. **21**, 1665-1677 (2009).

[14] P. Moetakef, T. A. Cain, D. G. Ouellette, J. Y. Zhang, D. O. Klenov, A. Janotti, C. G. V. d. Walle, S. Rajan, S. J. Allen, and S. Stemmer, *Electrostatic carrier doping of GdTiO$_3$/SrTiO$_3$ interfaces*, Appl. Phys. Lett. **99**, 232116 (2011).

[15] M. Imada, A. Fujimori, and Y. Tokura, *Metal-insulator transitions*, Rev. Mod. Phys. **70**, 1039-1263 (1998).

[16] Y. Tokura, Y. Taguchi, Y. Okada, Y. Fujishima, T. Arima, K. Kumagai, and Y. Iye, *Filling dependence of electronic properties on the verge of metal–Mott-insulator transition in Sr$_{1-x}$La$_x$TiO$_3$*, Phys. Rev. Lett. **70**, 2126-2129 (1993).

[17]  P. Moetakef, J. Y. Zhang, A. Kozhanov, B. Jalan, R. Seshadri, S. J. Allen, and S. Stemmer, *Transport in ferromagnetic GdTiO$_3$/SrTiO$_3$ heterostructures*, Appl. Phys. Lett. **98**, 112110 (2011).

[18]  B. Jalan, R. Engel-Herbert, N. J. Wright, and S. Stemmer, *Growth of high-quality SrTiO$_3$ films using a hybrid molecular beam epitaxy approach*, J. Vac. Sci. Technol. A **27**, 461-464 (2009).

[19]  J. Son, P. Moetakef, B. Jalan, O. Bierwagen, N. J. Wright, R. Engel-Herbert, and S. Stemmer, *Epitaxial SrTiO$_3$ films with electron mobilities exceeding 30,000 cm$^2$ V$^{-1}$ s$^{-1}$*, Nat. Mater. **9**, 482-484 (2010).

[20]  D. J. Keeble, B. Jalan, L. Ravelli, W. Egger, G. Kanda, and S. Stemmer, *Suppression of vacancy defects in epitaxial La-doped SrTiO$_3$ films*, Appl. Phys. Lett. **99**, 232905 (2011).

[21]  A. Ohtomo and H. Y. Hwang, *Surface depletion in doped SrTiO$_3$ thin films*, Appl. Phys. Lett. **84**, 1716-1718 (2004).

[22]  H. Suzuki, H. Bando, Y. Ootuka, I. H. Inoue, T. Yamamoto, K. Takahashi, and Y. Nishihara, *Superconductivity in single-crystalline Sr$_{1-x}$La$_x$TiO$_3$*, J. Phys. Soc. Jpn. **65**, 1529-1532 (1996).

[23]  M. Ziese, *Extrinsic magnetotransport phenomena in ferromagnetic oxides*, Rep. Prog. Phys. **65**, 143-249 (2002).

[24]   See supplemental material at [link to be inserted by publisher] for the resistance as a function of temperature and magnetoresistance data from a SrTiO$_3$/GdTiO$_3$/SrTiO$_3$ sample with a 0.8 nm thick quantum well.

[25]  H. D. Zhou and J. B. Goodenough, *Localized or itinerant TiO$_3$ electrons in RTiO$_3$ perovskites*, J. Phys.: Condens. Matter **17**, 7395–7406 (2005).

[26] J. F. Schooley, W. R. Hosler, E. Ambler, J. H. Becker, M. L. Cohen, and C. S. Koonce, *Dependence of the Superconducting Transition Temperature on Carrier Concentration in Semiconducting SrTiO₃*, Phys. Rev. Lett. **14**, 305-& (1965).

[27] B. Jalan and S. Stemmer, *Large Seebeck coefficients and thermoelectric power factor of La-doped SrTiO₃ thin films*, Appl. Phys. Lett. **97**, 042106 (2010).

[28] C. C. Hays, J. S. Zhou, J. T. Markert, and J. B. Goodenough, *Electronic transition in $La_{1-x}Sr_xTiO_3$*, Phys. Rev. B **60**, 10367-10373 (1999).

[29] Y. Okada, T. Arima, Y. Tokura, C. Murayama, and N. Mori, *Doping- and pressure-induced change of electrical and magnetic properties in the Mott-Hubbard insulator LaTiO₃*, Phys. Rev. B **48**, 9677-9683 (1993).

[30] M. Lee, J. R. Williams, S. Zhang, C. D. Frisbie, and D. Goldhaber-Gordon, *Electrolyte Gate-Controlled Kondo Effect in SrTiO₃*, Phys. Rev. Lett. **107**, 256601 (2011).

[31] J. Hubbard, *Electron Correlations in Narrow Energy Bands*, Proc. R. Soc. Lond. A **276**, 238-257 (1963).

[32] L. F. Mattheiss, *Effect of the 110°K Phase Transition on the SrTiO₃ Conduction Bands*, Phys. Rev. B **6**, 4740-4753 (1972).

[33] J. Appel, *Soft-Mode Superconductivity in $SrTiO_{3-x}$*, Phys. Rev. **180**, 508–516 (1969).

# Figure Captions

**Figure 1:** (a) Magnetoresistance of a 4 nm GdTiO$_3$/1 nm SrTiO$_3$/4 nm GdTiO$_3$/LSAT heterostructure at 15 K, 10 K, 5 K, and 2 K. Hysteresis appears below ~ 10 K in sweeps with increasing and decreasing B, respectively (see arrows). (b) Magnetoresistance of a 8 nm GdTiO$_3$/1 nm SrTiO$_3$/LSAT heterostructure at 5 K and 2 K.

**Figure 2:** Magnetization as a function of temperature at 10 mT for a 4 nm GdTiO$_3$/1 nm SrTiO$_3$/4 nm GdTiO$_3$/LSAT heterostructure. The dashed line is a guide to the eye. The inset shows the magnetization M as a function of magnetic field at 2 K.

**Figure 3.** (a) Magnetoresistance of La-doped SrTiO$_3$ film with a carrier density of $9\times10^{20}$ cm$^{-3}$ at 12 mK, showing hysteresis during sweeps with increasing and decreasing B, respectively.

**Figure 4.** (a) Magnetoresistance of a La-doped SrTiO$_3$ film with a carrier density of $1\times10^{20}$ cm$^{-3}$ as a function of temperature, measured with increasing B field. (b) Magnified magnetoresistance at 12 mK. The hysteresis in sweeps with increasing and decreasing B (see arrows) indicates the coexistence of superconductivity and ferromagnetism.

**Figure 5.** Superconducting and ferromagnetic transition temperatures of SrTiO$_3$ as a function of carrier concentration. Literature data for the superconducting transition temperatures is from Appel et al. (ref. [33]). The sheet carrier densities of the quantum wells ($n_{2D}$) were converted to 3D carrier concentrations ($n_{3D}$) according to $n_{3D} = n_{2D}/t \; [\text{cm}^{-3}]$, where $t$ is the thickness of the quantum well and $n_{2D}$ the carrier concentration determined from the Hall measurements at 5 K.

**Figure 1**

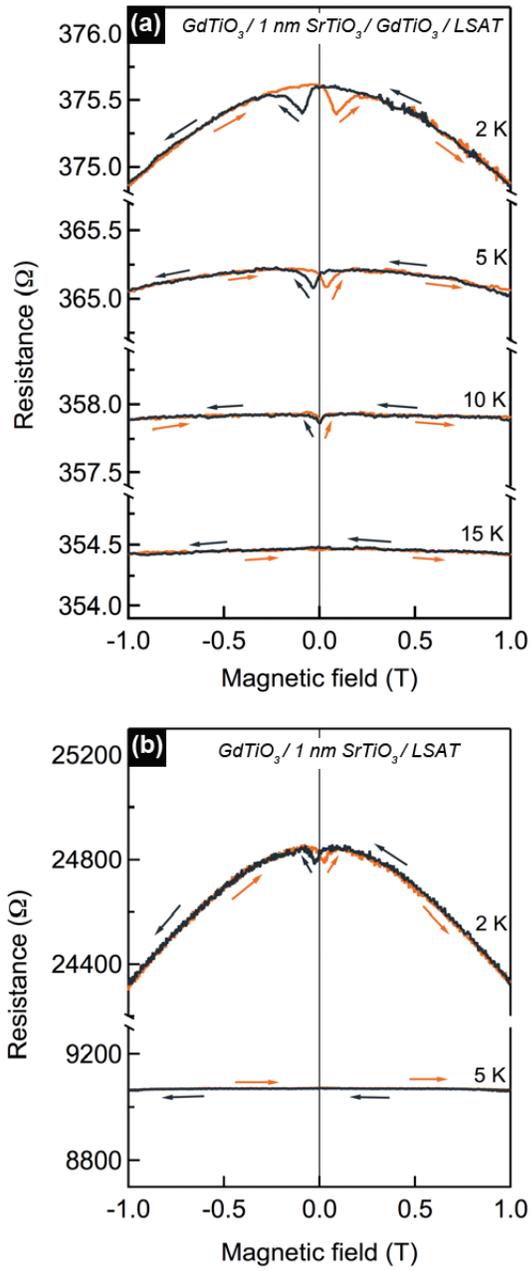

**Figure 2**

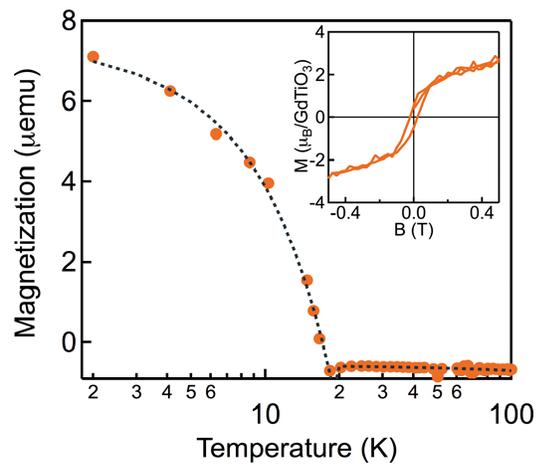

**Figure 3**

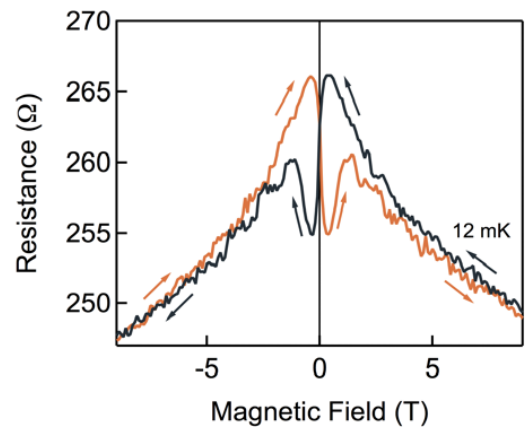

**Figure 4**

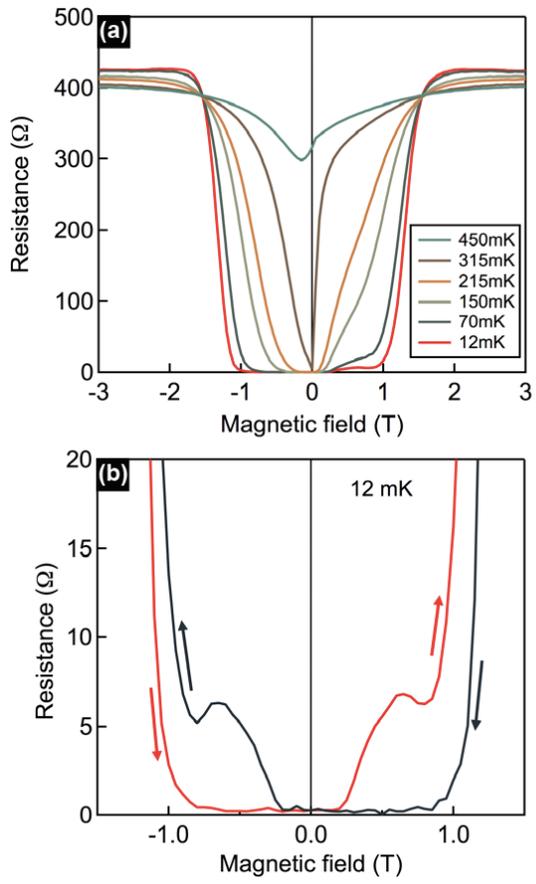

**Figure 5**

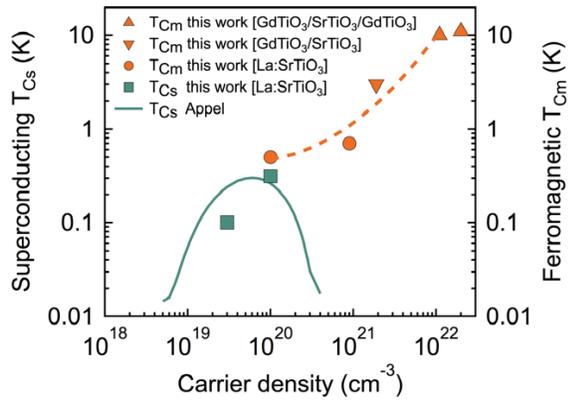

# Supplementary information: Carrier-controlled ferromagnetism in SrTiO$_3$

Pouya Moetakef[1], James R. Williams[2], Daniel G. Ouellette[3], Adam Kajdos[1], David Goldhaber-Gordon[2], S. James Allen[3], and Susanne Stemmer[1]

[1]Materials Department, University of California, Santa Barbara, California, 93106-5050, USA
[2]Department of Physics, Stanford University, Stanford, California, 94305-4045, USA
[3]Department of Physics, University of California, Santa Barbara, California, 93106-9530, USA

Figure S1 shows the magnetoresistance of a 4 nm GdTiO$_3$/0.8 nm SrTiO$_3$/4 nm GdTiO$_3$/LSAT heterostructure at temperatures between 12 and 2 K. The carrier concentration for this sample at room temperature is $8.22\times10^{14}$ cm$^{-2}$ and at 2 K it is $1.52\times10^{15}$ cm$^{-2}$, corresponding to a 3D carrier concentration of $1.9\times10^{22}$ cm$^{-3}$.

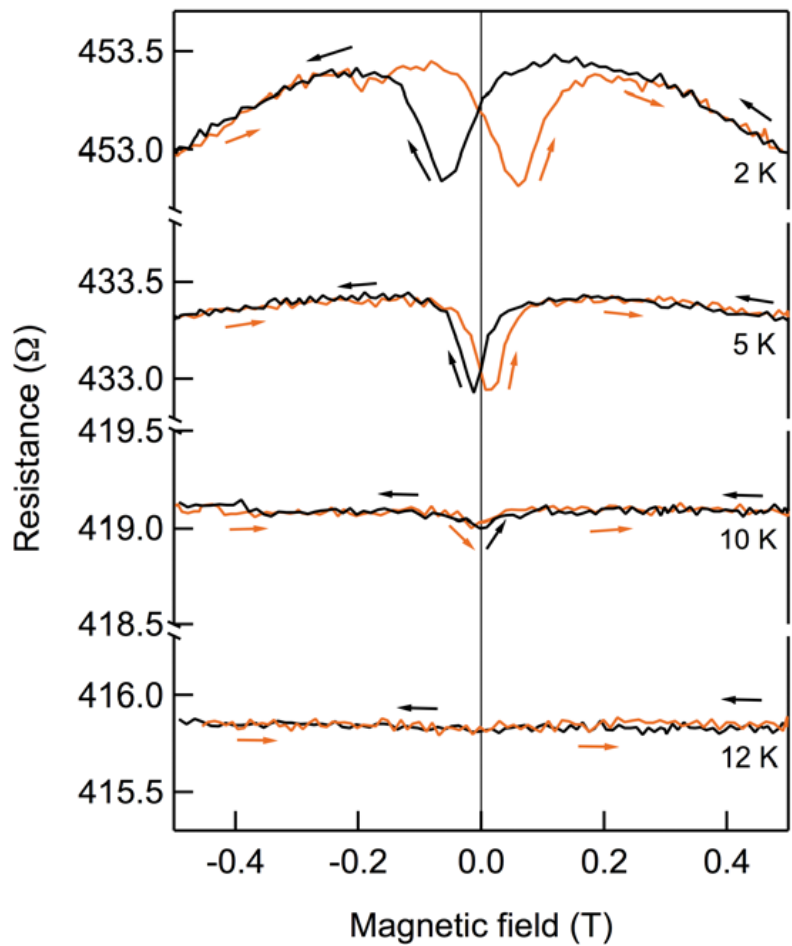

**Figure S1:** magnetoresistance of a 4 nm GdTiO$_3$/0.8 nm SrTiO$_3$/4 nm GdTiO$_3$/LSAT heterostructure at temperatures between 12 and 2 K. Hysteresis appears below ~ 10 K in sweeps with increasing and decreasing B, respectively (see arrows).

Figure S2 shows the sheet resistances and Hall carrier concentrations of the GdTiO$_3$/1-nm SrTiO$_3$/LSAT and GdTiO$_3$/1-nm SrTiO$_3$/ GdTiO$_3$/LSAT heterostructures.

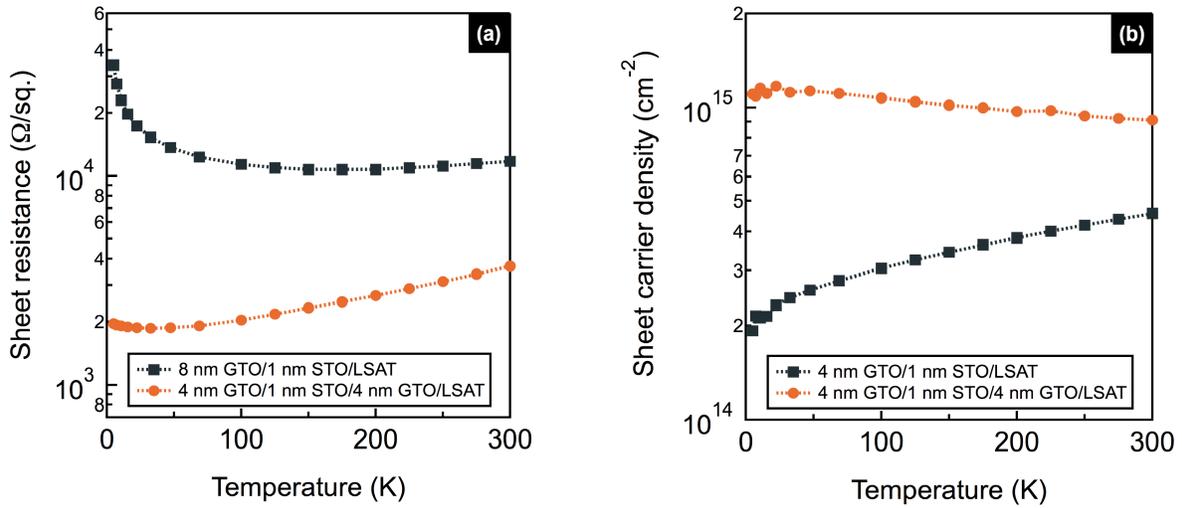

**Figure S2:** (a) Resistance and (b) carrier density as a function of temperature for two GdTiO$_3$/SrTiO$_3$ heterostructures.